# Optimizing Bio-energy Supply Chain to Achieve Alternative Energy Targets

[1]**Jubin Thomas**

[1]Independent Researcher, Media, Pennsylvania, USA, Email: jubin.thomas@ieee.org

## Abstract

In response to global warming and the dwindling reservoirs of fossil fuels, Thailand has increasingly embraced alternative energy sources. Central to its energy development strategy is the Alternative Energy Development Plan (AEDP), which aims to reduce energy intensity, capitalize on residual resources, and mitigate greenhouse gas emissions. While significant strides have been made in meeting various consumption targets set forth by the AEDP, notable challenges persist, particularly in the realms of bio-mass-derived electricity generation, bio-gas utilization, and bio-ethanol production from bio-mass. Therefore, this study delves into the factors contributing to the shortfall in achieving AEDP targets and proposes strategies to enhance the efficiency of the bio-energy supply chain. Leveraging mathematical and linear programming techniques, our research optimizes the supply chain dynamics, accounting for monthly supplier profiles spanning 77 provinces, 17 distinct biomasses, and 427 bio-energy plants equipped with diverse energy conversion technologies. Our findings indicate that Thailand currently boasts adequate bio-mass resources to fulfill the electricity and bio-ethanol targets outlined in the AEDP—provided enhancements are made to supply chain efficiency. To fully realize the objectives of the AEDP, we recommend augmenting bio-mass cultivation efforts and implementing new power plant installations. Additionally, we advocate for the consideration of high-methane content fuels, such as solid waste, as a means to alleviate bio-mass demand. This study underscores the paramount importance of strategic planning and optimization in propelling Thailand towards its alternative energy ambitions while surmounting supply chain impediments.

## 1. Introduction

In recent years, Thailand has been addressing energy and ecological challenges, spurred by rising energy costs, dwindling fossil fuel reserves, and environmental concerns, including natural disasters and greenhouse gas emissions. To mitigate these issues, there has been a growing interest in renewable energy sources due to their lower carbon emissions and eco-friendliness [1]–[3]. Projections indicate a significant increase in the adoption of renewable energy, potentially supplanting fossil fuels as the primary energy source. In line with this vision, Thailand has actively engaged in various regional and international initiatives, such as the Asia-Pacific Economic Cooperation (APEC), ASEAN Economic Community (AEC), and the Paris Agreement. As part of its commitments, Thailand has pledged to reduce energy intensity per Gross Domestic Product (GDP) by 30% by 2036 compared to 2010 levels, in accordance with APEC guidelines. Additionally, under the ASEAN framework, Thailand has committed to promoting renewable energy trade among member countries. Domestically, Thailand has devised the "Thailand Integrated Energy Blueprint (TIEB)", overseen by the Ministry of Energy (MOE), to guide its energy development endeavors. This blueprint undergoes periodic review and refinement to align with Thailand's evolving economic, security, and environmental priorities. The TIEB comprises five interconnected plans—the Power Development Plan (PDP), Energy Efficiency Development Plan (EEDP or EEP), Alternative Energy Development Plan (AEDP), Oil Plan, and Gas Plan, each tailored to specific objectives. However, the AEDP—for instance, outlines renewable energy deployment targets and consumption ratios over a 20-year horizon. While progress has been made in realizing alternative energy goals, some targets remain unmet, as evidenced by the disparities between actual and targeted consumption in 2021. Notably, electricity generation from small hydropower, bio-mass, bio-gas, solid waste, and bio-ethanol falls short of the 2021 targets, with varying degrees of shortfall ranging from 11.14% to 34.46%. In contrast, wind power and bio-diesel have surpassed their respective consumption targets. Despite the shortfall in specific sectors, Thailand has achieved its overall alternative energy production target, with wind power compensating for deficits in other areas. However, the root causes hindering the realization of the year-2021 AEDP goal in actual electricity consumption derived from bio-mass, bio-gas (from energy crops), and bio-ethanol. Therefore, utilizing the Python programming language, we employ mathematical programming techniques to simulate the logistical operations within Thailand's bio-energy supply chain. Our analytical framework encompasses bio-mass requirements, production capacities, and logistics flow throughout the supply chain. Through simulation, we aim to understand the current state of the bio-energy supply chain and identify areas for improvement to align with the AEDP targets. Our objectives are as follows:

- Develop a mathematical model and simulate Thailand's bio-energy supply chain, encompassing bio-mass, bio-gas (from energy crops), and ethanol production.
- Investigate the underlying issues contributing to the shortfall in real bio-energy consumption compared to the AEDP target for the year 2021.
- Provide recommendations to enhance Thailand's bio-energy supply chain to meet the AEDP objectives by 2021.

While various alternative energy sources contribute to Thailand's energy landscape, our focus is specifically on electricity generated from bio-mass, bio-gas from energy crops, and bio-ethanol production from cassava and molasses. It's important to note that our study excludes small hydropower, bio-gas from non-energy-crop sources, and solid waste. Bio-mass feedstocks originate solely from within Thailand, sourced from the country's six major economic crops—rice, sugarcane, corn, cassava, oil palm, and coconut, resulting in a total of 17 bio-mass types considered in our analysis. Therefore, in our optimization approach, we assume retailers can distribute all purchased electricity from bio-energy power plants to end consumers, without accounting for limitations in electrical transmission lines.

The paper is as follows; related studies are shown in the following section. The methods and materials are detailed in Section 3. The experimental analysis is carried out in Section 4, and in Section 5, we provide some conclusions and plans for future research.

## 2. Related Works

Thailand, located in Southeast Asia, exports diverse agricultural products including rice, sugarcane, corn, cassava, rubber trees, oil palm, and coconuts. These crops generate significant bio-mass residues, such as rice husks, bagasse, corn leaves, cassava rhizomes, oil palm bunches, and coconut shells, which offer potential for sustainable energy production according to [4]. Despite this, the full utilization of agricultural bio-mass remains largely untapped, highlighting opportunities for both energy generation and waste management. Thailand's industrial heat energy, as detailed in [5], derives significantly from renewable sources within the agricultural sector, including sugar, palm oil, cassava starch, processed wood, rice mills, and livestock farming. Primarily, the sugar industry harnesses heat from bagasse and rice husk. According to [6]—projections in the PDP anticipate a final energy consumption of 182,720 ktoe by 2036, while the EEDP targets a 30% reduction in energy intensity compared to 2010 levels, aiming for a final consumption of 131,000 ktoe by 2036. Additionally, the AEDP aims for 20% of electricity production to stem from alternative sources by 2036. Therefore, to facilitate alternative energy production, the Thailand government has transitioned from an adder system to a Feed-in-Tariff (FiT) proposed by [7]—incentivizing private sector investment in renewable energy projects. However, challenges persist, notably in the capacity of existing electrical transmission infrastructure to accommodate both traditional and alternative energy sources at maximum production levels. Consequently, initiatives are underway to expand transmission line capacity, as evaluated by the Electricity Generating Authorization of Thailand (EGAT).

Numerous studies have investigated the optimization of bio-energy supply chains, particularly focusing on bio-fuel supply chains such as [8]. An overview and critical analysis of bio-fuel supply chain optimization were conducted by [8], providing insights into logistical activities and the varying objectives, decision levels, and methodologies associated with this optimization process. Their study primarily concentrated on the utilization of forest and wood chip bio-mass. Additionally, [9] developed a mathematical model for optimizing bio-fuel supply chains, specifically targeting cellulosic bio-ethanol production, while also addressing concerns related to land usage change, environmental and social impacts, and governmental policies. Employing Mixed Integer Linear Programming (MILP), [10] optimized grass-based bio-ethanol supply chains with different harvesting methods, aiming to minimize total annual costs and determine optimal facility locations. Furthermore, [11] utilized Arena Simulation Software (ASS) to simulate bio-fuel supply chains, focusing on minimizing feedstock delivery costs, energy consumption, and greenhouse gas emissions. Their simulation model provided comprehensive economic and environmental performance evaluations under various conditions. Additionally, [12] explored the impact of feedstock quality on overall costs and supply chain modeling, proposing a two-stage stochastic modeling approach to optimize different scenarios by considering moisture content fluctuations. Bio-mass and bio-gas supply chains share fundamental concepts. The prevalence of bio-mass supply chains is particularly notable in Asia, with a significant focus in Thailand. A comprehensive review by [13] examines various bio-mass feedstocks across different regions and their associated energy conversion technologies. Specifically, [14] undertook a case study in Thailand, employing MILP to model the bio-mass supply chain. Their objective was to minimize the overall cost of the supply chain while considering factors such as bio-mass perishability. Moreover, [15] proposed a methodology to formulate fuel supply strategies and evaluate associated risks affecting bio-mass quality and pricing, alongside assessing bio-mass potential in northeast Thailand.

## 3. Materials and Methods

To conduct our study, we first identify supply chain participants and activities. Next, we create a mathematical model reflecting these activities and supply chain context. Therefore, we gather necessary data for model inputs and apply linear programming to optimize performance metrics and decision variables. Finally, we analyze outcomes and iteratively refine the model. See Fig. 1 for a visual representation of our methodology.

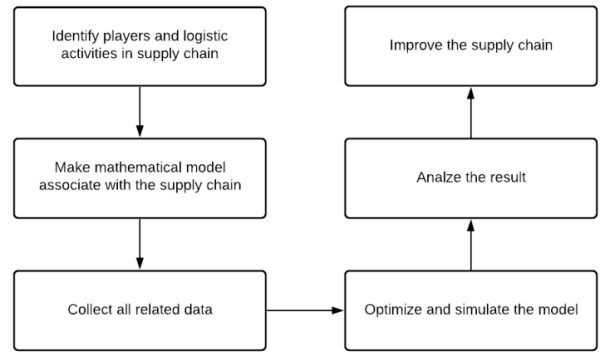

Fig. 1. Workflow for operations

However, a notable challenge arises from the lack of precise data regarding individual consumer bio-ethanol demand [16]–[28]. Therefore, we opt to consider only the aggregate bio-ethanol production. Our model comprises 77 suppliers, each representing a distinct province, supplying varied bio-mass types and quantities. The prevalent mode of transportation in Thailand, our focal region, is the 10-wheel truck. The production technologies employed by power plants encompass direct-firing, co-generation, or gasification, with each plant specializing in one of these methods. Meanwhile, bio-ethanol plants utilize an anaerobic process for production. It is important to reiterate that retailers are regarded as consumers within our model framework, as previously indicated. The current study examines an annual timeframe, divided into months, focusing on the operations of 77 suppliers supplying varying quantities of 17 different types of bio-masses. Our analysis encompasses a total of 427 bio-energy plants, comprising 215 bio-mass power plants, 185 bio-gas power plants, and 27 bio-ethanol plants. These plants utilize five distinct energy production technologies. Specifically, in bio-mass power plants, energy production involves direct-firing, gasification, and co-generation techniques, denoted by $q = 1, 2$, and $3$ respectively. In bio-gas power plants, anaerobic digestion ($q = 4$) is employed, while bio-ethanol plants utilize the fermentation process ($q = 5$). Our analysis concentrates on the dynamics of bio-mass supply chains for energy production, considering variables such as bio-mass quantity from suppliers, utilization in energy production, and plant inventory capacities. Parameters crucial to supply chain efficiency include bio-mass availability, supplier-plant distances, and plant characteristics. Therefore, we explore plant energy production capacity, bio-mass-to-energy conversion factors, efficiency, capacity factors, and financial costs to sustain the supply chain. The annual bio-mass cost is calculated as the product of the bio-mass cost and the quantity of bio-mass delivered to the plants. Similarly, the annual transportation cost involves the product of the transportation variable cost, the quantity of bio-mass delivered to the plants, and the distance between suppliers and plants. The annual operational cost is divided into three categories—electricity production from bio-mass, electricity production from bio-gas, and bio-ethanol production. Each category's cost is determined by the respective operational cost of the power plants and the electricity or bio-ethanol produced. The holding cost is computed as the product of the holding cost per unit of bio-mass and the inventory level. Constraints are imposed on the logistic activities in the supply chain. These constraints limit the availability of bio-mass for delivery to the plants. Additionally, each plant has a maximum inventory level that cannot be exceeded. After bio-mass is delivered to the plants and stored in inventory, it is used as a feedstock for energy production. The change in inventory level from one time-step to another is determined by the amount of bio-mass received from suppliers and the bio-mass used for energy production. It is ensured that the feedstocks withdrawn from inventory do not exceed the inventory level. Each plant has a maximum energy production capacity. The energy production constraints are established to ensure that the energy produced does not exceed these limits. Furthermore, the energy produced must meet the national demands. The electricity produced from bio-mass and bio-gas must match the corresponding national electricity demands. Similarly, the bio-ethanol produced must meet the national bio-ethanol demand. Finally, all variables are constrained to be non-negative.

### 3.1 Dataset Analysis

Comprehensive data collection is essential for our optimization model, covering suppliers, power plants, bio-ethanol facilities, and consumers at each time-step. Bio-mass and energy plant data are sourced from Department of Alternative Energy Development and Efficiency (DEDE[1]) and MOE[2], with bio-mass diversity and crop abundance from Office of Agricultural Economics (OAE[3]). Bio-mass availability is quantified using crop quantities and ratio coefficients, with bio-mass prices and heat capacities provided in Tables 1 and 2. Information on methane components and heat equivalents is presented in Table 3, with methane density reported at 0.657 kg/m$^3$ and specific heat capacities listed as $c_v$ = 27.4 J/mol.K or 1.709 kJ/kg.K, and $c_p$ = 35.8 J/mol.K or 2.232 kJ/kg.K. In bio-ethanol production, molasses and peeled cassava are the sole bio-mass sources, with conversion coefficients of 3.8 kg/liter and 5.975 kg/liter of bio-ethanol, respectively. Geographical coordinates from Google Maps and Python-Open Source Routing Machine (OSRM[4]) package are used to calculate distances between supply chain components. Transportation relies on a 10-wheel truck for bio-mass and petroleum tank trucks for molasses, with variable costs including fuel, tire, and maintenance expenses, totaling 6.561 Thai Baht (THB) per kilometer. Bio-mass transportation cost per kilometer per ton depends on the truck's loading capacity per trip, considering bio-mass densities outlined in Table 4—to determine whether weight or cargo dimensions limit the bio-mass weight per truck. The approach to calculating variable transportation costs for bio-mass involves several steps:

1. Determine the 10-wheel truck's variable cost in THB per kilometer.
2. Calculate the maximum bio-mass weight of the truck can load by comparing cargo dimensions and weight capacity limitations.
3. Divide Step 1 by Step 2 to obtain the variable transportation cost in THB per kilometer per ton.

The calculated variable bio-mass transportation costs are summarized in Table 5. Plant operational costs are assessed through Levelized Cost of Energy (LCOE), covering investment, operation, maintenance, and feedstock expenses, denoted in USD per kWh. Investment LCOE ranges from 0.04-0.13 USD/kWh for direct-firing, 0.11-0.28 USD/kWh for gasification, and 0.07-0.29 USD/kWh for co-generation technologies.

Table 1. Cost of green biomass

| Biomass | Price (THB/ton) |
| --- | --- |
| rice straw | 2000 |
| rice husk | 1500 |
| sugarcane leaves | 800 |
| bagasse | 500 |
| molasses | 10,410 |
| corn leaves and tops | 1500 |
| corncob | 500 |
| peeled cassava | 2700 |
| cassava rhizome | 1800 |
| cassava fiber | 3300 |
| cassava peels | 2800 |
| oil palm bunch | 50 |
| oil palm fiber | 1500 |
| oil palm shell | 3200 |
| coconut bunch | 1000 |
| coconut bract | 5000 |
| coconut shell | 1000 |

Table 2. A list of crops and the biomass they produce

| Biomass | Heat Capacity (MJ/ton) |
| --- | --- |
| rice straw | 12,330 |
| rice husk | 13,520 |
| sugarcane leaves | 15,480 |
| bagasse | 7,370 |
| molasses | No usage in heat conversion. Use only for bioethanol conversion. |
| corn leaves and tops | 9,830 |
| corncob | 9,620 |
| peeled cassava | No usage in heat conversion. Use only for bioethanol conversion. |
| cassava rhizome | 5,490 |
| cassava fiber | 1,470 |
| cassava peels | 1,490 |
| oil palm bunch | 7,240 |
| oil palm fiber | 11,400 |
| oil palm shell | 16,900 |
| coconut bunch | 15,400 |
| coconut bract | 16,230 |
| coconut shell | 17,930 |

Table 3. Methane obtained from biomass

| Biomass | Methane Content (m³/kg) | Heat Equivalent (MJ/ton) |
|---|---|---|
| rice straw | 0.226 | 108.0402 |
| rice husk | 0.019 | 9.0830 |
| sugarcane leaves | 0.148 | 70.7520 |
| bagasse | 0.185 | 88.4400 |
| molasses | 0.324 | No usage in biogas conversion. Use only for bioethanol conversion. |
| corn leaves and tops | 0.199 | 95.1327 |
| corncob | 0.1 | 47.8054 |
| peeled cassava | 0.262 | No usage in biogas conversion. Use only for bioethanol conversion. |
| cassava rhizome | 0.09676 | 46.2565 |
| cassava fiber | 0.167 | 79.8350 |
| cassava peels | 0.078 | 37.2882 |
| oil palm bunch | 0.1996 | 95.4196 |
| oil palm fiber | 0.1664 | 79.5482 |
| oil palm shell | No usage in biogas conversion. | 0 |
| coconut bunch | No usage in biogas conversion. | 0 |
| coconut bract | No usage in biogas conversion. | 0 |
| coconut shell | No usage in biogas conversion. | 0 |

Table 4. Densities of biomass

| Biomass | Density (kg/m³) |
|---|---|
| rice straw | 178.255 |
| rice husk | 356.065 |
| sugarcane leaves | 190.43 |
| bagasse | 160 |
| molasses | 1441 |
| corn leaves and tops | 81.61 |
| corncob | 182.38 |
| peeled cassava | 637.38 |
| cassava rhizome | 238 |
| cassava fiber | 712.50 |
| cassava peels | 247.87 |
| oil palm bunch | 380 |
| oil palm fiber | 250 |
| oil palm shell | 400 |
| coconut bunch | 355 |
| coconut bract | 151.91 |
| coconut shell | 920.53 |

Table 5. Variable cost of transportation

| Biomass | Variable Cost (THB/km.ton) |
|---|---|
| rice straw | 1.01 |
| rice husk | 0.51 |
| sugarcane leaves | 0.95 |
| bagasse | 1.13 |
| molasses | 0.23 |
| corn leaves and tops | 2.21 |
| corncob | 0.99 |
| peeled cassava | 0.41 |
| cassava rhizome | 0.76 |
| cassava fiber | 0.41 |
| cassava peels | 0.73 |
| oil palm bunch | 0.47 |
| oil palm fiber | 0.72 |
| oil palm shell | 0.45 |
| coconut bunch | 0.51 |
| coconut bract | 1.19 |
| coconut shell | 0.41 |

## 4. Experimental Analysis

In our pursuit to align bio-mass and plant profiles with the targeted demand outlined in the AEDP, it became evident that the existing resources fell short. Therefore, the primary inquiry arose—what is the maximum bioenergy potential achievable in Thailand? To address this question, we redefined three crucial national consumption parameters—electricity demand from biomass ($D_{B,Elc}$), electricity demand for biomass-generated electricity ($D_{G,Elc}$), and bioethanol demand ($D_{Eth}$), transforming them into variables. Simultaneously, we adjusted our objective function to maximize annual energy output. These variables were constrained within the thresholds of targeted consumptions—3940 MW for $D_{B,Elc}$, 387 MW for $D_{G,Elc}$, and 4.79 ML/day for $D_{Eth}$. Standardizing energy output in MWh, we computed the product of $D_{B,Elc}$ and $D_{G,Elc}$ by 24, while for $D_{Eth}$, considering that one liter of bioethanol yields energy equivalent to $5.9313 \times 10^{-3}$ MWh, we multiplied it by the bioethanol demand variable, summing them across all time-steps. The optimization process took 4,564 seconds to complete, revealing Thailand's bioenergy production potential. Our findings indicate a capacity of up to 33.02 TWh, distributed across electricity generated from biomass (2,550.4 MW), biogas (35.65 MW), and bioethanol (4.79 ML/day) Therefore, we conducted an analysis, labeled as "case 1", to assess the behavior of the supply chain at the current maximum bioenergy potential as shown in Figs. 2, 3, and 4. The optimization process lasted 2,378 seconds, resulting in a total cost of 375.22 billion THB. Notably, certain biomass types remained unused, while others were utilized at varying capacities. In this scenario, the capacity factor of all plants was observed to be 0.89, with biomass power plants operating at full capacity. However, despite efforts to meet demands, a significant portion of biogas power plants remained inactive. Inventory management revealed fluctuations, particularly during periods of low biomass availability.

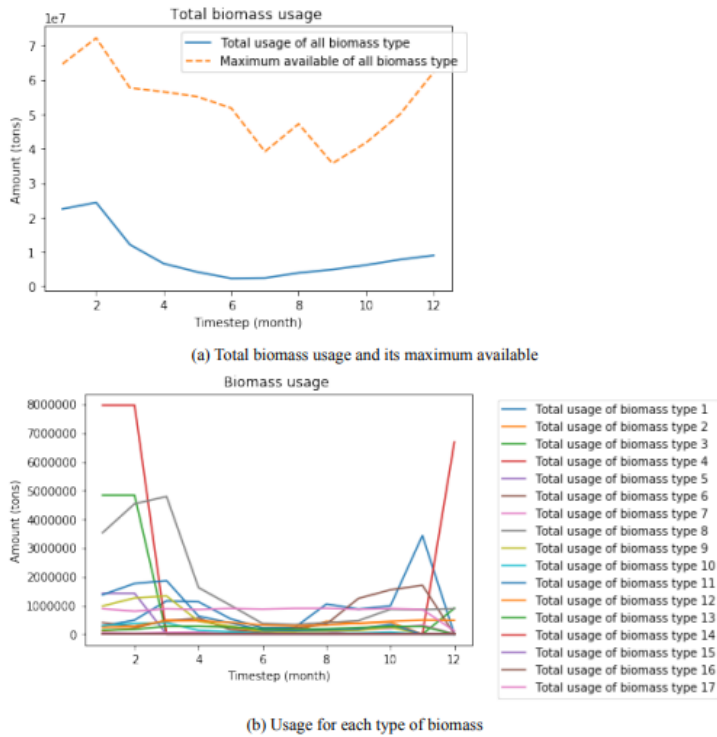

(a) Total biomass usage and its maximum available

(b) Usage for each type of biomass

Fig. 1. Case 1: Use of biomass

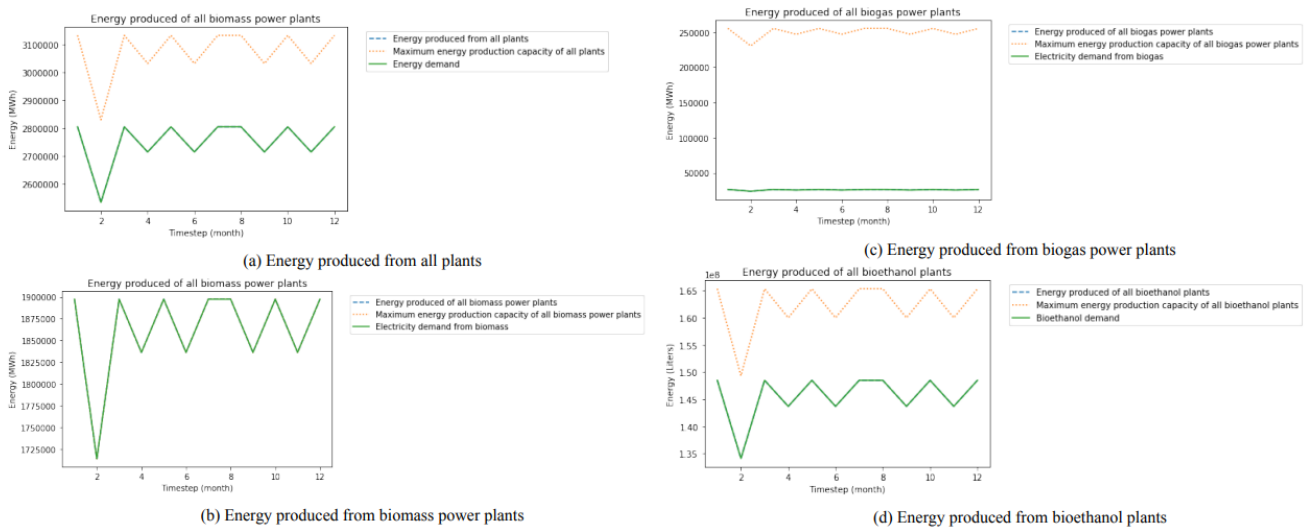

(a) Energy produced from all plants

(c) Energy produced from biogas power plants

(b) Energy produced from biomass power plants

(d) Energy produced from bioethanol plants

Fig. 2. Case 1: Plant-based energy production

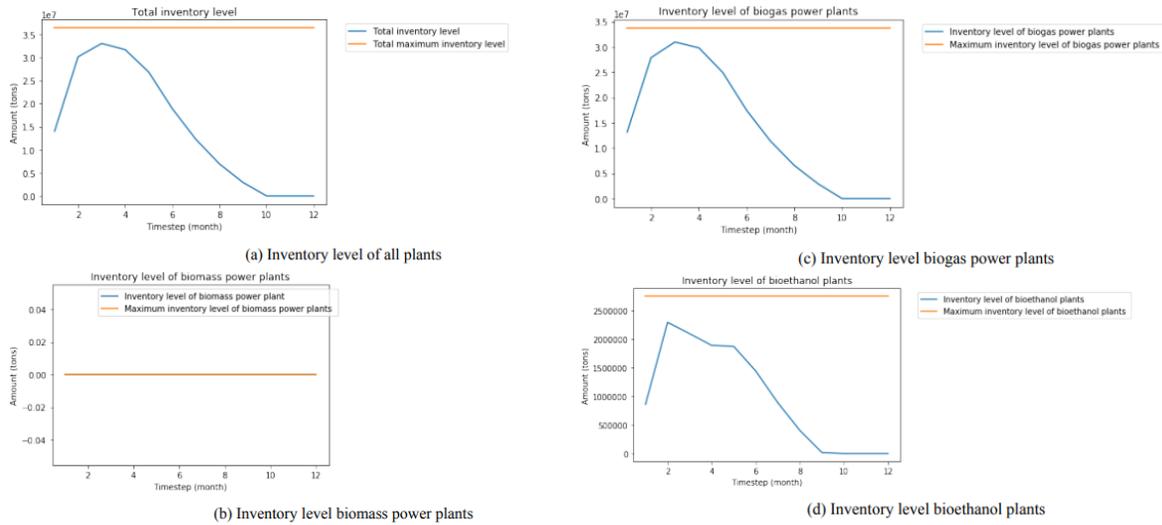

(a) Inventory level of all plants       (c) Inventory level biogas power plants

(b) Inventory level biomass power plants       (d) Inventory level bioethanol plants

Fig. 3. Case 1: Plant inventory level

Subsequently, in "case 2", we aimed to maximize plant operation and minimize inventory variance as shown in Figs. 4, 5, and 6. The optimization process lasted 8,613 seconds, resulting in a total cost of 499.99 billion THB. Similar to the previous case, specific biomass types were underutilized. In both cases, biomass power plants operated at full capacity, while bioethanol plants contributed significantly to meeting demand. Notably, all biogas power plants were operational in "case 2", contrasting with the previous scenario. Inventory management improvements led to reduced fluctuations and increased overall utilization. Despite these enhancements, challenges remain in fully optimizing plant operation and minimizing inventory variance. The results of case 1 and case 2 comparisons reveal a considerable discrepancy in total costs, with case 1 exhibiting a lower total cost of 124.77 billion THB compared to case 2. In case 1, biomass material cost, transportation cost, and holding cost are respectively 33.94 billion THB, 69.9 billion THB, and 20.87 billion THB lower than in case 2. It's noteworthy that operational costs for biomass power plants, biogas power plants, and bioethanol plants remain constant, as they generate equivalent energy output in both cases. The parity in energy production across both cases ensures an unaltered overall energy output, thus maintaining the capacity factor at a consistent level, as depicted in Fig. 7. However, when evaluating individual plant operations, case 2 demonstrates 100% operational efficiency across all plants, in contrast to 75.87% in case 1. Specifically, while biomass power plants and bioethanol plants operate at full capacity in both cases, only 44.32% of biogas power plants are operational in case 1. Comparatively, the overall inventory utilization in case 2 surpasses that of case 1 by 30.32%, with biomass power plants and bioethanol plants registering 91.27% and 19.79% higher utilization rates, respectively. In contrast, biogas power plants in case 2 record a slightly lower utilization rate of 0.41% compared to case 1. Regarding inventory level variance, the overall populational variance of all plants in case 1 exceeds that of case 2 by $6.519 \times 10^{13}$ tons$^2$. However, it's notable that while the populational variance of biomass power plants remains constant, biogas power plants exhibit a higher variance of $3.6 \times 10^{11}$ tons$^2$ compared to case 2, whereas bioethanol plants show a lower variance of $6.14 \times 10^{10}$ tons$^2$.

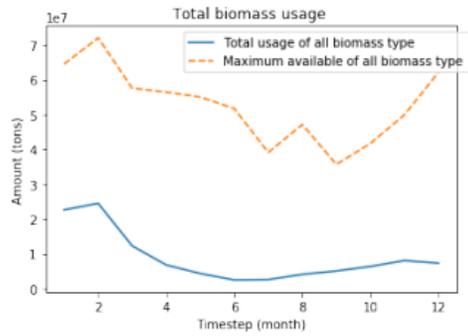

(a) Total biomass usage and its maximum available

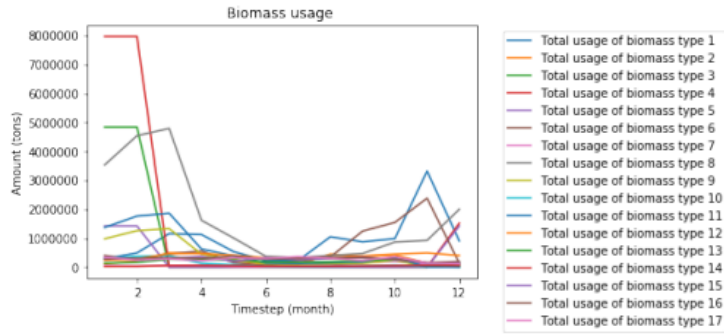

(b) Usage for each type of biomass

Fig. 4. Case 2: Use of biomass

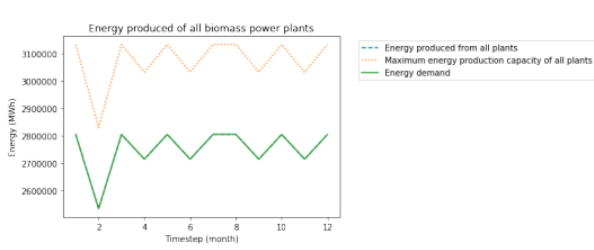

(a) Energy produced from all plants

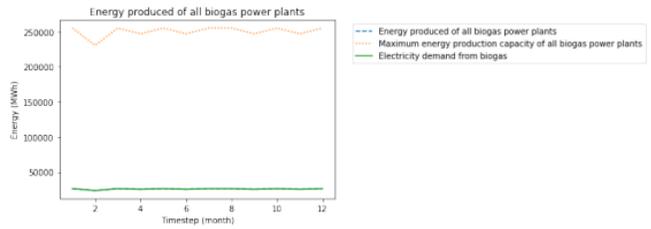

(c) Energy produced from biogas power plants

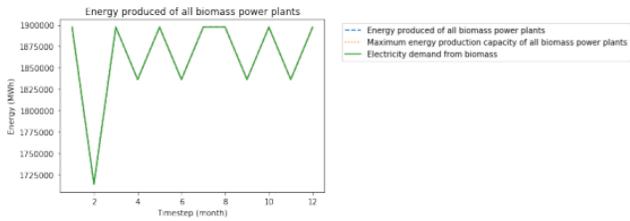

(b) Energy produced from biomass power plants

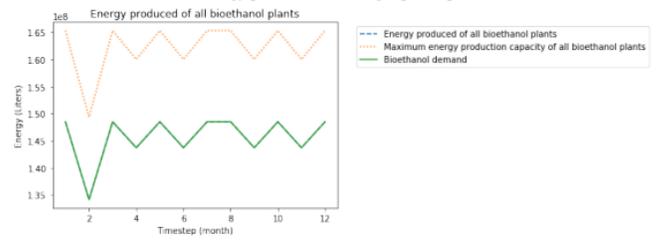

(d) Energy produced from bioethanol plants

Fig. 5. Case 2: Plant-based energy production

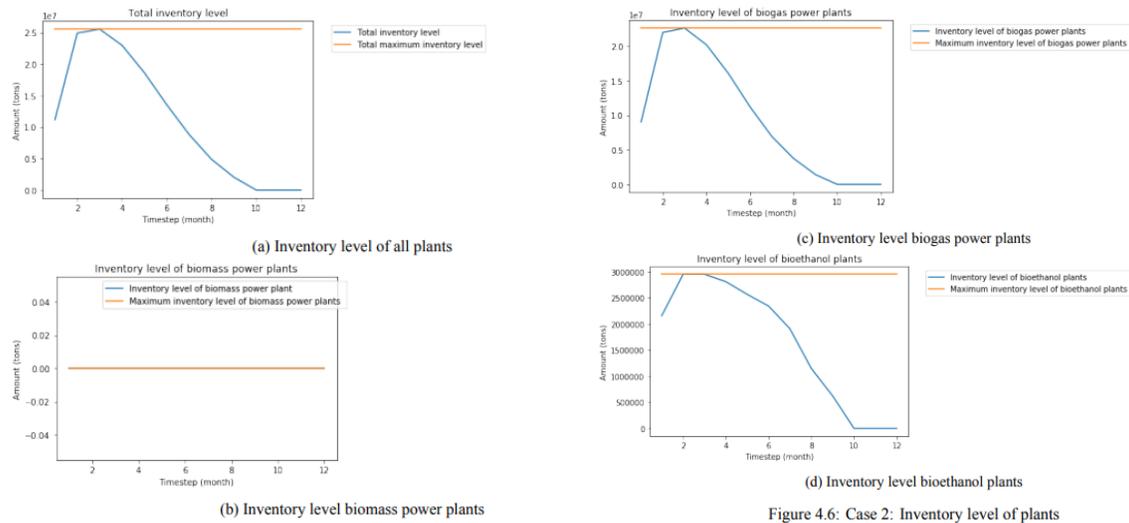

(a) Inventory level of all plants

(b) Inventory level biomass power plants

(c) Inventory level biogas power plants

(d) Inventory level bioethanol plants

Figure 4.6: Case 2: Inventory level of plants

Fig. 6. Case 2: Plant inventory level

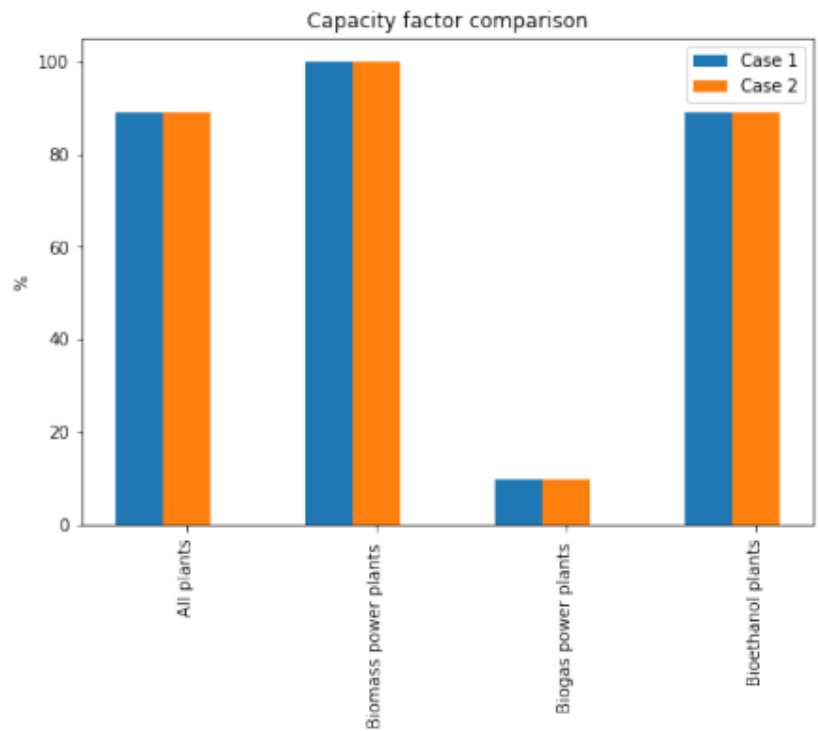

Fig. 7. Comparison of case 1 and case 2's capacity factors

## 4.1 Additional Experiments

The current bio-energy potential falls short of meeting these targets, barring bio-ethanol production. Despite the theoretical maximum capacities of biomass and biogas power plants being 3,000.47 MW and 403.50 MW respectively, accounting for maintenance reduces the effective capacity to 342.97 MW for biomass and remains unaltered for biogas. This shortfall necessitates the establishment of additional biomass and biogas power plants, along with increased cultivation of energy crops. Thus, key questions arise regarding the requisite crop cultivation areas and quantities, as well as the installation capacities of new power plants to fulfill the targets. To address these queries, we introduce novel parameters such as the index of new power plants, additional biomass, and the maximum production capacity of new biomass power plants. To ascertain the additional biomass needed, we reformulate the objective function to minimize this requirement while accommodating the new variables. Though the essence of the constraints remains akin to prior sections, adjustments are made to accommodate the introduction of new variables. Considering the technology employed, we presume all new biomass power plants utilize gasification, while biogas plants persist with anaerobic digestion. Constraints regarding biomass availability are augmented to include deliveries to both existing and new plants, ensuring compliance with supply limitations.

Inventory and production constraints for new biomass and biogas power plants necessitate additional formulations, maintaining the balance between operational requirements and inventory management. Furthermore, energy production constraints are refined to accommodate the introduction of new power plants and align with demand expectations. Upon optimization, our analysis reveals a substantial requirement for additional molasses, equivalent to 4.65 billion tons of sugarcane harvesting. This indicates a need for significant expansion in sugarcane cultivation. Additionally, considerations are made for by-products such as sugarcane leaves, trunks, and bagasse. Detailed results, including biomass supplier profiles and the impact of increased crop cultivation, are provided for comprehensive understanding in Figs. 8, 9, and 10. The optimal location for new power plants is determined through the center of gravity method, minimizing supply chain costs while factoring in transportation distances. Finally, we address the determination of maximum production capacities for new power plants, alongside the broader implications for the supply chain.

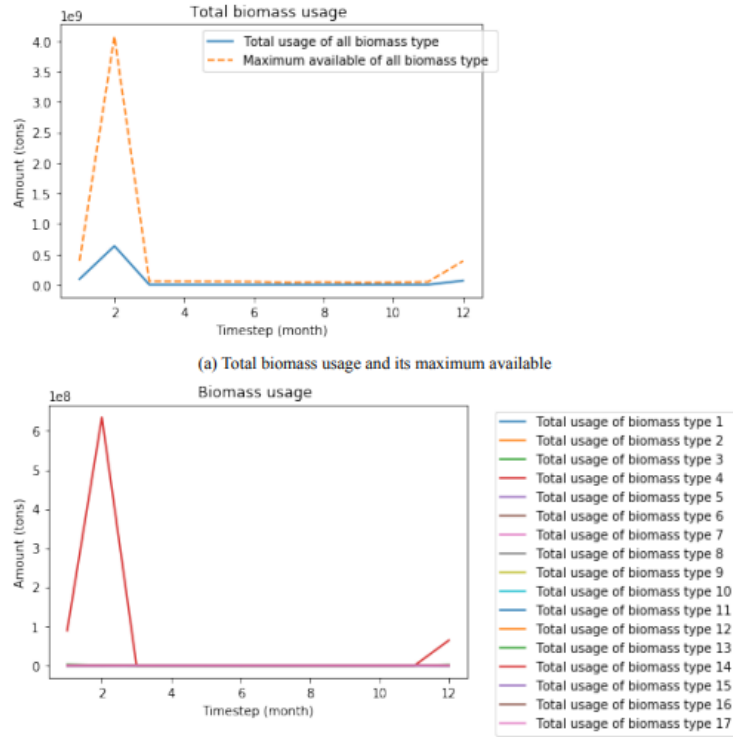

Fig. 8. Case 3: Use of biomass

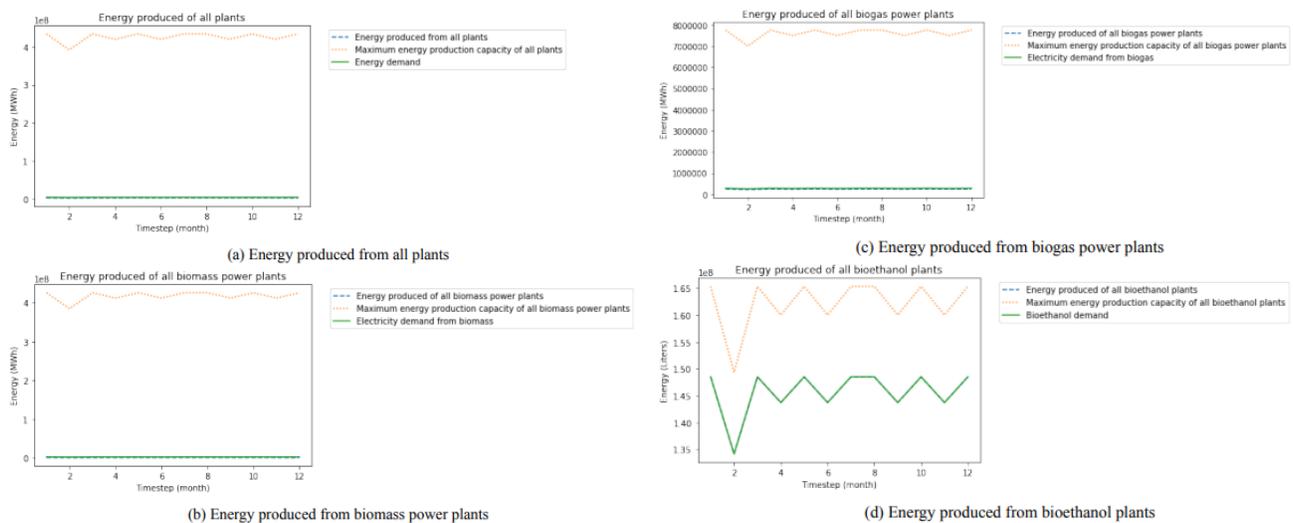

Fig. 9. Case 3: Plant-based energy production

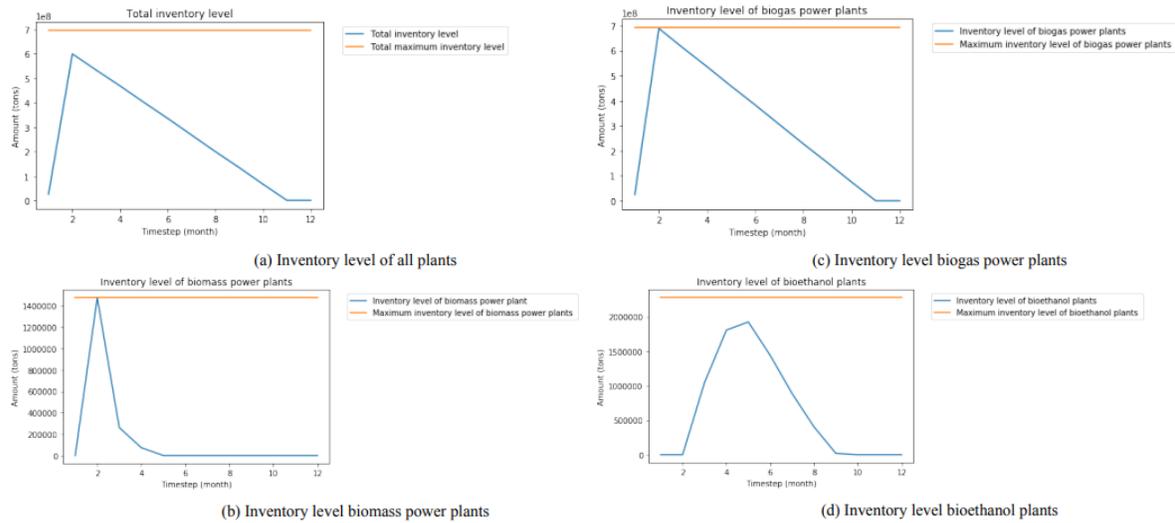

(a) Inventory level of all plants      (c) Inventory level biogas power plants

(b) Inventory level biomass power plants      (d) Inventory level bioethanol plants

Fig. 10. Case 3: Plant inventory level

## 5. Conclusion and Future Works

In this study, we utilized Python programming with the PuLP package to conduct linear programming optimization tasks for the bioenergy supply chain. We analyzed data from 77 provinces, considering 17 biomasses and 427 bioenergy plants employing five energy conversion technologies. Our study delved into three cases—evaluating Thailand's bioenergy potential (Case 1), optimizing supply chain operations (Case 2), and addressing biomass requirements and new power plant locations (Cases 3). Case 1 showcased Thailand's capacity to meet AEDP targets with 4.79 million litres of bioethanol daily and substantial electricity generation from biomass and biogas. Case 2's show supply chain cost optimization yielded a cost of 375.22 billion THB, accompanied by specific operational metrics. Case 2 minimized maximum inventory levels, resulting in a 33.25% cost increase from Case 1 to 499.99 billion THB. Despite this, it achieved full plant operation (100%) with improved supply chain utilization. Cases 3 emphasized the trade-off between cost and facility utilization, with identifying sugarcane requirements and optimal plant locations. While favouring bio-mass sources with higher methane content like solid waste or dung can enhance electricity generation, this study exclusively used linear programming. Employing advanced techniques such as non-linear programming or supply chain simulation software could yield more optimal outcomes yet faced challenges due to optimization package limitations with large datasets. Moreover, addressing electrical transmission line constraints, as discussed in above sections, is crucial for enhancing the systematicity and realism of supply chain optimization.

023-01721-W.